# Performance Evaluation of Multimedia Traffic in Cloud Storage Services over Wi-Fi and LTE Networks

Albert Espinal[1], V. Sanchez Padilla[2,3], and
Yesenia Cevallos[4]

[1] Escuela Superior Politécnica del Litoral, ESPOL, Telematics Engineering Dept., Km 30.5 Via Perimetral, Guayaquil, EC090902, Ecuador
aespinal@espol.edu.ec
[2] Virginia Polytechnic Institute and State University, Engineering Education Dept., Blacksburg, VA 24061, United States
[3] Universidad ECOTEC, College of Engineering, Architecture, and Natural Sciences Km 13.5 Samborondón, Samborondón, EC092302, Ecuador
[4] Universidad Nacional del Chimborazo, College of Engineering, Km 1.5 Vía Guano, Riobamba, Ecuador

**Abstract.** The performance of Dropbox, Google Drive, and OneDrive cloud storage services was evaluated under Wi-Fi and LTE network conditions during multimedia file uploads. Traffic was captured using Wireshark, and key metrics (including delay, jitter, bandwidth, and packet loss) were analyzed. Google Drive maintained the most consistent performance across both types of networks, showing low latency and reduced jitter. Dropbox showed efficient bandwidth utilization, but experienced a longer delay over LTE, attributed to a greater number of intermediate hops. OneDrive presented variable behavior, with elevated packet rates and increased sensitivity to fluctuations in the mobile network. A bimodal distribution of packet sizes was observed and modeled using a dual Poisson function. In general, Wi-Fi connections provided greater stability for multimedia transfers, while LTE performance varied depending on platform-specific implementations. The results contribute to a better understanding of traffic behavior in cloud-based storage applications and suggest further analysis with larger datasets and heterogeneous access networks.

## 1 Introduction

Cloud-based storage services like Google Drive, Dropbox, and OneDrive have become essential tools for managing and sharing digital content across several settings [1, 2]. These platforms are now crucial for storage, upload, and stream of multimedia files, such as videos, images, and audio, on several devices and networks [3, 4]. The increasing reliance on cloud storage for multimedia content has resulted in substantial data traffic, especially as users demand seamless access and real-time synchronization of large media files across different network environments [5, 6], including Wi-Fi and mobile LTE networks.

Unlike traditional streaming services, which often rely on direct server-to-user delivery, cloud storage platforms handle asynchronous file uploads, downloads, and real-time access to data [7, 8]. This introduces a variety of network performance challenges, particularly related to factors like latency, jitter, and packet loss [9, 10]. As a result, the performance of cloud-based storage applications in handling multimedia traffic is directly impacted by the underlying network infrastructure, which can vary widely between Wi-Fi and mobile environments [11, 12]. Evaluating the performance of these platforms under different conditions is critical to understanding their efficiency and stability.

Multimedia content, such as high-definition video, audio, and images, has seen rapid growth, driven by the widespread use of streaming services, social media, and video conferencing tools [13, 14]. This growth has driven a change in the way multimedia content is accessed and consumed, in which users expect high-quality and low-latency experiences, regardless of device, scenario, or location [15,16]. In this context, the need for efficient data management and real-time delivery has highlighted the importance of reliable network performance. For instance, projections estimate that Netflix, one of the most popular streaming platforms, will reach close to 300 million subscribers by 2029, showing a constant growth trend [17].

The demand for multimedia content has led to an increased need for advanced technological solutions to ensure smooth user experiences [18, 19]. As applications rely on high-bandwidth and steady networks to manage large volumes of data, aspects such as access speed, data consistency, and real-time handling of multimedia traffic become crucial. To evaluate these factors, performance metrics such as delay, jitter, packet loss, and available bandwidth must be carefully considered [20, 21].

Although multimedia content continues to grow, most of the studies so far have focused on how it moves across networks, without a focused look at what happens on specific cloud-based storage platforms. There is still a lack of research that connects multimedia traffic with the actual performance of cloud services. The literature on this topic can be limited to showing how delay, jitter, or packet loss affect the user experience when uploading or downloading media files on these specific platforms. The motivation for this study is to analyze how cloud-based storage services handle multimedia traffic under different network conditions and to identify how performance metrics influence the overall experience.

## 2   State of the Art

The performance of multimedia traffic in cloud storage applications has changed in recent years due to the growing demand for digital content and ongoing improvements in network infrastructure. Cloud storage remains essential for han-dling multimedia content due to its scalability, global accessibility, and lower infrastructure costs [22, 23]. Providers like Cloudflare, Akamai, and Amazon CloudFront have strengthened their networks to improve multimedia delivery. These platforms enable geographically distributed access, which reduces latency and improves the end user experience [24].

The deployments of 5G and early stage 6G developments have changed the way multimedia traffic (e.g., virtual reality, video streaming, real-time gaming) is delivered. These technologies support ultra-high data rates, low delay, and more bandwidth [25, 26]. When combined with cloud storage platforms, they allow efficient real-time multimedia delivery, even in ultra HD, 4K, or 8K formats [27]. Decentralized storage systems, such as IPFS (InterPlanetary File System), also help improve content availability and reduce delay through distributed data location [28]. In addition, Content Delivery Networks (CDN) still play a key role in optimizing multimedia traffic. By placing servers in multiple geographic regions, CDNs reduce the physical distance between users and content, accelerating the delivery, and reducing the network load [24].

CDN vendors, such as Akamai or Cloudflare, have upgraded their platforms with AI-based load balancing and traffic prediction. These tools dynamically adjust resources according to user demand and network conditions [29]. This helps to minimize packet loss and jitter, and supports consistent real-time media delivery. Moreover, modern video codecs like H.266/VVC (Versatile Video Coding) and AV1 became more widely used in recent years. They reduce bandwidth usage without sacrificing video quality [30, 31], which is especially important in cloud-based applications that manage large multimedia files. Furthermore, adaptive bitrate streaming adjusts the quality of the media in real time based on current network performance [32].

Cloud platforms have integrated AI and machine learning to manage multimedia traffic more efficiently as well. Predictive algorithms analyze usage patterns to prevent congestion and balance load. Several of these platforms use AI to automatically allocate resources in data centers, which improves scalability and reduces access time [33]. These systems also detect and respond to quality issues related to packet loss or delay [34]. However, a key factor that can affect multimedia traffic performance, cloud storage infrastructure, and service availability is the presence of cyberattacks. There are investigations that analyze how Distributed Denial-of-Service (DDoS) attacks affect network systems and servers, where researchers suggest mitigation strategies to reduce their impact [35].

## 3 Methodology

This study shows the performance evaluation of three cloud storage services: Dropbox, Google Drive, and One Drive, which can be easilty configured by end users [2]. These platforms were selected because they are widely used and popular [36, 37]. The experiments were set up to measure key performance metrics such as delay, jitter, bandwidth, and packet loss during file transfers under different network conditions. Figure 1 shows the topology configured to analyze and assess cloud-based storage services. A laptop represents the customer, uploading and downloading files to and from each evaluated platform. The computer was a Lenovo laptop with a Core i3 processor, 8GB of RAM, 64-bit CPU, and Windows 11 features, to mention a few.

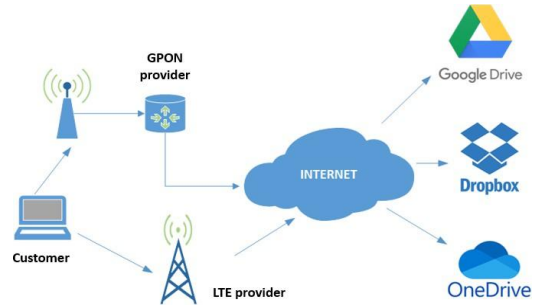

**Fig. 1.** Block diagram.

The test used multimedia files ranging from 35 to 40 MB to simulate typical user workloads. File transfers were carried out over two types of networks: a Wi-Fi connection with an average speed of 50 Mbps, representing a controlled home setup, and an LTE mobile network averaging 25 Mbps, which showed more variability in connection quality. They ran each network-platform combination three times to ensure consistent and reliable results.

The monitoring process relied on Wireshark [38] to capture and analyze data network traffic. The analysis focused on TCP traffic over IPv4, using specific filters to track packets such as TCP retransmissions and Transport Layer Security (TLS) protocol exchanges. Input/output traffic graphs were generated to identify patterns, such as traffic peaks and idle periods during upload sessions.

The experiment involved the uploading of files from the customer to each selected cloud storage platform. During each session, real-time metrics were recorded, and performance differences between both types of networks were noticed. This approach helped to evaluate how well each platform handled multimedia transfers under varying connectivity conditions. The packet lengths were also analyzed and a bimodal traffic pattern was found, which confirms observations from previous work carried out by Espinal et al. [39]. In addition, the present study focused on IPv4 traffic, as it remains the most widely implemented. The test environment used for data collection did not support end-to-end IPv6 connectivity.

## 4   Result Analysis

The Dropbox tests using the Wi-Fi connection recorded an exchange of 33,120 packets over 84 seconds. This resulted in an average rate of 392 packets per second, an average packet size of 1,275 bytes, and a transfer rate of 4 Mbps. Figure 2 shows this bandwidth usage pattern. The entire transfer was performed over the IPv4 protocol, with 100% of the application packets encapsulated in TCP. At the same time, 33.2% of that traffic used the TLS protocol. The packet size distribution showed that 42.89% fell in the small range of 50 to 99 bytes, while 57.04% were large packets between 1,450 and 1,500 bytes. Figure 3 illustrates this distribution.

The traffic captured on port 443 (HTTPS) confirmed secure upload and download operations. The relevant packet types appeared throughout the session, including ACK, SYN, FIN, and application-layer data packets. These packets confirmed efficient file transfer flows, with idle periods between peaks that pointed to server-side acknowledgments to guarantee data integrity. Retransmissions remained at a minimal level, suggesting a stable and secure connection. This behavior reflects the solid design and ability of Dropbox to handle file transfers efficiently over Wi-Fi.

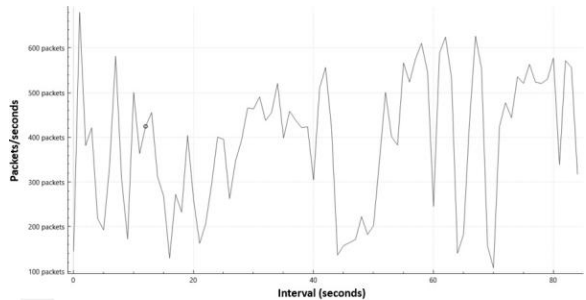

**Fig. 2.** Bandwidth consumption, Dropbox on Wi-Fi network

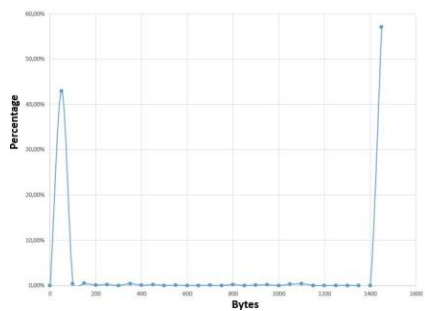

**Fig. 3.** Packet frequency by size on Dropbox

Based on the data analyzed in the previous section, several traffic models were estimated using the Poisson probability distribution function, applied to IPv4 traffic. The model was calculated for both the wireless network and the LTE access network, since the traffic trends were similar in both cases. The adjusted model for total traffic on the wireless network uses a combination of two Poisson distributions, with parameters $\lambda_1 = 64.30$ and $\lambda_2 = 1424.00$. The probability that a packet length follows the first distribution is 0.430. For the second distribution, the probability reaches 0.570. Equation (1) shows the model, which is the result of the sum of the two Poisson distributions.

$$P(X = x) = 0.430 \cdot \frac{e^{-64.30} \cdot 64.30^x}{x!} + 0.570 \cdot \frac{e^{-1466} \cdot 1466^x}{x!} \quad (1)$$

For the LTE transfer case, 29.54% of the packets had an average size of 58 bytes, while 70.21% averaged 1,314 bytes. PingPlotter was used to measure delay, jitter, and packet loss during the Dropbox test. The average delay reached approximately 88.2 milliseconds, the jitter measured 15.12 milliseconds, and the packet loss remained close to 1%, as shown in Figures 4 and 5. The route to the destination included a total of 17 hops, which added to the overall delay observed in the test.

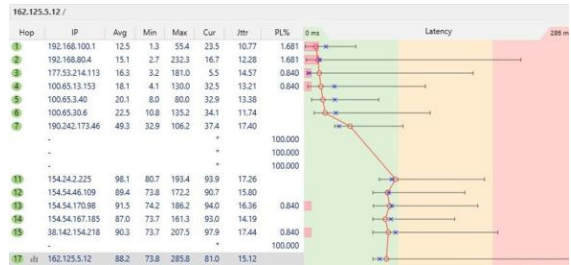

**Fig. 4.** Traffic metrics collected for Dropbox

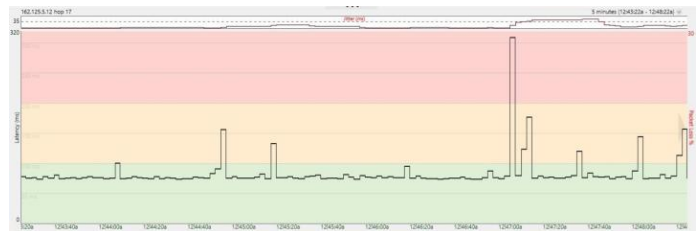

**Fig. 5.** Parameter measurement during test interval for Dropbox

TLS traffic over port 443 indicates secure data exchange with Dropbox servers. However, TCP retransmissions were observed. This likely resulted from congestion or instability in the local network, which increases latency and reduces system efficiency.

For the Google Drive scenario, the transfer time was 70 seconds. During this period, 45,136 packets were exchanged, with an average packet size of 945 bytes, an average transfer rate of 637 packets per second, and an average bandwidth usage of 4.813 Mbps, as shown in Figure 6. The transfer process occurred entirely over IPv4, where 100% of the application packets were encapsulated in TCP. In addition, 6.3% of the packets used the TLS protocol. A smaller portion of the packets was associated with the Google QUIC protocol.

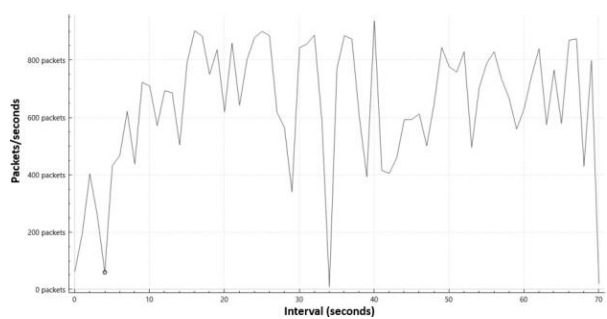

**Fig. 6.** Bandwidth consumption, Google Drive on Wi-Fi network

The packet length analysis showed a bimodal pattern. Approximately 35.36% of the packets fell within the small size range of 50 to 99 bytes, while 60.93% were large packets between 1,450 and 1,500 bytes. The corresponding Poisson model for this traffic appears in equation (2). During the LTE connection, 41.25% of the packets averaged 58 bytes, while 57.90% showed an average size of 1,314 bytes. The average delay was around 27.7 milliseconds, jitter reached 11.16 milliseconds, and the packet loss remained close to 5%, as shown in Figure 7. The route to the destination included 11 hops.

$$P(X = x) = 0.370 \cdot \frac{e^{-56.74} \cdot 56.74^x}{x!} + 0.630 \cdot \frac{e^{-1466} \cdot 1466^x}{x!} \qquad (2)$$

For the Microsoft OneDrive scenario, the transfer took 96 seconds. During that period, 54,028 packets were exchanged, with an average packet size of 803 bytes, an average rate of 557 packets per second, and an average bandwidth usage of 3.578 Mbps, as shown in Figure 8. Furthermore, traffic directed to the domain 1drv.ms (Microsoft) and encrypted with TLS 1.3 showed significant variation in data volume. Transfer peaks reached up to 800 KB/s, alternating with periods of inactivity. High retransmission rates during specific moments suggest packet loss or temporary congestion. Although the TCP protocol handles retransmissions, performance decreases in terms of latency and overall efficiency.

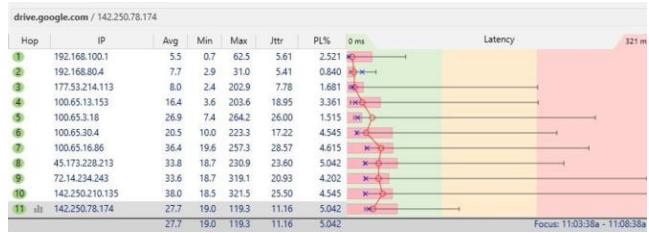

**Fig. 7.** Traffic metrics collected for Google Drive

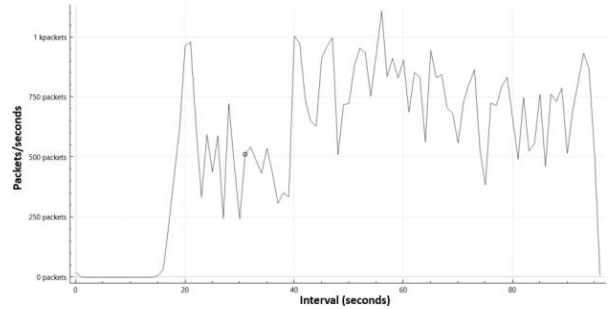

**Fig. 8.** Bandwidth consumption, One Drive on Wi-Fi network

The transfer process took place entirely over the IPv4 protocol, with 100% of the application packets encapsulated in TCP. Of that traffic, 5.7% used the TLS protocol. The analyzed packet lengths showed a bimodal distribution, where 45.66% of the packets fell in the range of 50 to 99 bytes (small packets), whereas 52.19% were large packets between 1,450 and 1,500 bytes. The Poisson model for this traffic appears in equation (3). For the case of LTE, 42.13% of the packets averaged 58 bytes, whereas 57.72% averaged 1,314 bytes.

$$P(X = x) = 0.465 \cdot \frac{e^{-56.76} \cdot 56.76^x}{x!} + 0.535 \cdot \frac{e^{-1466} \cdot 1466^x}{x!} \qquad (3)$$

The average delay reached approximately 104.9 milliseconds. The maximum jitter peaked at 9.45 milliseconds, and packet loss remained around 5%, as shown in Figure 9. The path to the destination included a total of 24 hops, which directly impacted the measured delay.

The traffic capture between the local IP and an external server on port 443 shows instances of TCP ACKs and out-of-order packets, caused by fluctuations in mobile connection quality. The mobile network introduces delays and packet loss, leading to instability and performance variations. Compared to the Wi-Fi connection, the average traffic volume is lower, but the packet sizes remain similar on the different platforms evaluated in this study. Table 1 summarizes the main metric results for the cloud storage platforms over the Wi-Fi connection. Table 2 presents the results collected for LTE connectivity.

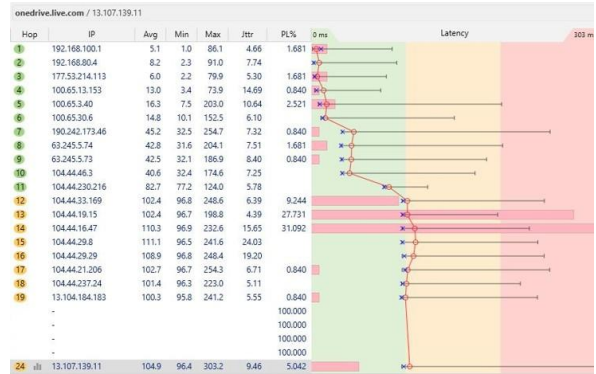

**Fig. 9.** Traffic metrics collected for One Drive

**Table 1.** Comparison of Cloud Storage Service Metrics over Wi-Fi Network

| Parameter | Dropbox | Google Drive | One Drive |
|---|---|---|---|
| Total Data Transferred (MB) | 43 | 44 | 45 |
| Transfer Time (s) | 84 | 70 | 96 |
| Packets Transferred | 33120 | 45136 | 54028 |
| Packet Rate (packets/s) | 392 | 637 | 557 |
| Average Packet Size (bytes) | 1275 | 945 | 803 |
| Bandwidth (Mbps) | 3.997 | 4.813 | 3.578 |
| Average Delay (ms) | 88.2 | 27.7 | 104.9 |
| Jitter (ms) | 15.12 | 11.6 | 9.6 |
| Packet Loss (%) | 1.2 | 5 | 5 |
| Number of Hops | 17 | 11 | 24 |

**Table 2.** Comparison of Cloud Storage Service Metrics over LTE Network

| Parameter | Dropbox | Google Drive | One Drive |
|---|---|---|---|
| Total Data Transferred (MB) | 44 | 44 | 45 |
| Transfer Time (s) | 41 | 91 | 33 |
| Packets Transferred | 45212 | 54370 | 55079 |
| Packet Rate (packets/s) | 1086 | 591 | 1643 |
| Average Packet Size (bytes) | 942 | 791 | 786 |
| Bandwidth (Mbps) | 8.182 | 3.742 | 10.700 |
| Average Delay (ms) | 154.2 | 58.9 | 135.3 |
| Jitter (ms) | 33.94 | 10.17 | 9.34 |
| Packet Loss (%) | 0.84 | 3 | 0.84 |
| Number of Hops | 10 | 9 | 29 |

## 5    Conclusions

The performance analysis of cloud storage services under varying network conditions revealed key differences in efficiency and stability between platforms. Wi-Fi networks offered a more reliable environment for cloud storage transfers. In mobile networks, platform-specific factors, such as how each service handles retransmission and buffering, played a critical role in performance. In the case of Dropbox, this platform showed efficient bandwidth use by transferring fewer bytes and packets, along with larger average packet sizes and lower packet loss, indicating stable transfers over both Wi-Fi and LTE. However, its delay and jitter increased significantly in mobile networks, likely due to a higher number of end-to-end hops.

For the case of Google Drive, it achieved the best overall performance, offering a robust option to manage segmentation and retransmission efficiently, even under fluctuating conditions. In Wi-Fi, it maintained regular transfer peaks with low delay and jitter. In LTE, it preserved acceptable performance, probably influenced by fewer intermediate hops. On the other hand, OneDrive exhibited more variable behavior. It recorded the longest transfer times on Wi-Fi and the shortest on LTE. High delay and elevated packet loss aligned with its higher hop count. This platform also generated the highest packet rate due to smaller packet sizes, and showed greater sensitivity to unstable network conditions.

The study faced a few limitations. Variability in mobile network quality may have influenced results and file sizes were limited to 40 MB, limiting extrapolation to high-volume scenarios. Despite this, the experiments successfully captured the relative behavior of each platform under controlled and comparable conditions. Future work could extend this analysis to larger datasets, different mobile carriers, or edge-computing contexts, to better understand how cloud performance scales in more demanding use cases.

## References

1. T. Tsalko and S. Nevmerzhytska, "Cloud technologies: use in the educational process as a way to high management in business," *Zeszyty Naukowe Wyżsszej Szko-ly Ekonomiczno-Spo-lecznej w Ostro-lece*, vol. 50, no. 3, pp. 1–12, 2023.
2. A. S. Ogunmodede and E. O. Adepoju, "Cloud computing: A novel digital storage paradigm," *Journal of Economic, Social and Educational*, vol. 2, no. 2, pp. 108–116, 2023.
3. A. A. Laghari, X. Zhang, Z. A. Shaikh, A. Khan, V. V. Estrela, and S. Izadi, "A review on quality of experience (QoE) in cloud computing," *Journal of Reliable Intelligent Environments*, vol. 10, no. 2, pp. 107–121, 2024.
4. M. L. Rathod, A. Shivaputra, H. Umadevi, K. Nagamani, and S. Periyasamy, "Cloud computing and networking for SmartFarm AgriTech," *Journal of Nanomaterials*, vol. 2022, no. 1, p. 6491747, 2022.
5. M. Mokale, "Best practices for real-time data processing in media applications," *International Journal of Multidisciplinary Research and Growth Evaluation*, vol. 1, no. 5, pp. 131–137, 2020.

6. S. Khriji, Y. Benbelgacem, R. Ch´eour, D. E. Houssaini, and O. Kanoun, "Design and implementation of a cloud-based event-driven architecture for real-time data processing in wireless sensor networks," *The Journal of Supercomputing*, vol. 78, no. 3, pp. 3374–3401, 2022.
7. A. Alam, "Cloud-based e-learning: scaffolding the environment for adaptive e-learning ecosystem based on cloud computing infrastructure," in *Computer Communication, Networking and IoT: Proceedings of 5th ICICC 2021, Volume 2*, pp. 1–9. Springer, 2022.
8. S. Holloway, "The impact of cloud-based information systems on collaboration and productivity in remote teams," *Preprints.org*, 2024.
9. S. M. D. S. AHMAD, "CHALLENGES IN TRANSMISSION OF DATA AND NETWORK CONNECTIVITY," *Authorea Preprints*, 2023.
10. A. Q. Khan, M. Matskin, R. Prodan, C. Bussler, D. Roman, and A. Soylu, "Cloud storage cost: a taxonomy and survey," *World Wide Web*, vol. 27, no. 4, p. 36, 2024.
11. W. A. Aziz, "Performance Evaluation of Voice over WiFi (VoWiFi) Using IP Multimedia Subsystem (IMS)," *International Journal of Simulation–Systems, Science & Technology*, vol. 24, no. 2, 2023.
12. S. K. Gopal, A. S. Mohammed, V. R. Saddi, N. Jiwani, and J. Logeshwaran, "Exploring the Quality of Service Impacts of Cloud Computing over Wireless Networks," in *2024 International Conference on E-mobility, Power Control and Smart Systems (ICEMPS)*, 2024, pp. 01–06. doi:10.1109/ICEMPS60684.2024.10559341.
13. L. Zhang-Kennedy and S. Chiasson, "A systematic review of multimedia tools for cybersecurity awareness and education," *ACM Computing Surveys (CSUR)*, vol. 54, no. 1, pp. 1–39, 2021.
14. I. Chatterjee and P. Chakraborty, "Use of information communication technology by medical educators amid COVID-19 pandemic and beyond," *Journal of Educational Technology Systems*, vol. 49, no. 3, pp. 310–324, 2021.
15. W. Moina-Rivera, M. Garcia-Pineda, J. Guti´errez-Aguado, and J. M. Alcaraz-Calero, "Cloud media video encoding: review and challenges," *Multimedia Tools and Applications*, vol. 83, no. 34, pp. 81231–81278, 2024.
16. W. Wang, X. Wei, W. Tao, M. Zhou, and J. Cheng, "Quality of Experience-Oriented Cloud-Edge Dynamic Adaptive Streaming: Recent Advances, Challenges, and Opportunities," *Symmetry*, vol. 17, no. 2, p. 194, 2025.
17. O. Zagorodnia, "Influence of Product Placement in Netflix Original Shows on Consumers' Brand Recall," Master's thesis, Webster University, 2024.
18. M. T. Sultan and H. El Sayed, "QoE-Aware Analysis and Management of Multimedia Services in 5G and Beyond Heterogeneous Networks," *IEEE Access*, vol. 11, pp. 77679–77688, 2023. doi:10.1109/ACCESS.2023.3298556.
19. X. He and Q. Zhang, "Cloud Computing Based Digital Media Content Distribution Technology," *Procedia Computer Science*, vol. 247, pp. 461–468, 2024. The 11th International Conference on Applications and Techniques in Cyber Intelligence. doi:10.1016/j.procs.2024.10.055.
20. A. Ali and A. Ware, "Effective performance metrics for multimedia mission-critical communication systems," *Annals of Emerging Technologies in Computing (AETiC)*, vol. 5, no. 2, pp. 1–14, 2021.
21. D. Chefrour, "One-way delay measurement from traditional networks to SDN: A survey," *ACM Computing Surveys (CSUR)*, vol. 54, no. 7, pp. 1–35, 2021.
22. Y. Mansouri, A. N. Toosi, and R. Buyya, "Data storage management in cloud environments: Taxonomy, survey, and future directions," *ACM Computing Surveys (CSUR)*, vol. 50, no. 6, pp. 1–51, 2017.

23. M. Liu, L. Pan, and S. Liu, "Cost optimization for cloud storage from user perspectives: Recent advances, taxonomy, and survey," *ACM Computing Surveys*, vol. 55, no. 13s, pp. 1–37, 2023.
24. H. Bayat Pour, "Web Content Delivery Optimization," Master's thesis, Aalto University, 2016.
25. F. Salahdine, T. Han, and N. Zhang, "5G, 6G, and Beyond: Recent advances and future challenges," *Annals of Telecommunications*, vol. 78, no. 9, pp. 525–549, 2023.
26. M. Bozanic and S. Sinha, *Mobile communication networks: 5G and a vision of 6G*. Springer, 2021.
27. T. Janevski, *Future Fixed and Mobile Broadband Internet, Clouds, and IoT/AI*. John Wiley & Sons, 2024.
28. T. V. Doan, Y. Psaras, J. Ott, and V. Bajpai, "Toward Decentralized Cloud Storage With IPFS: Opportunities, Challenges, and Future Considerations," *IEEE Internet Computing*, vol. 26, no. 6, pp. 7–15, 2022. doi:10.1109/MIC.2022.3209804.
29. S. Shetty, T. M. S., V. H. M., and R. N. Shaikh, "Intelligent Network Traffic Control with AI and Machine Learning," in *2024 IEEE 16th International Conference on Computational Intelligence and Communication Networks (CICN)*, 2024, pp. 353–357. doi:10.1109/CICN63059.2024.10847397.
30. M. Lee, H. Song, J. Park, B. Jeon, J. Kang, J.-G. Kim, Y.-L. Lee, J.-W. Kang, and D. Sim, "Overview of versatile video coding (H. 266/VVC) and its coding performance analysis," *IEIE Transactions on Smart Processing & Computing*, vol. 12, no. 2, pp. 122–154, 2023.
31. A. Punchihewa and D. Bailey, "A Review of Emerging Video Codecs: Challenges and Opportunities," in *2020 35th International Conference on Image and Vision Computing New Zealand (IVCNZ)*, 2020, pp. 1–6. doi:10.1109/IVCNZ51579.2020.9290536.
32. G. Gao and Y. Wen, "Video transcoding for adaptive bitrate streaming over edge-cloud continuum," *Digital Communications and Networks*, vol. 7, no. 4, pp. 598–604, 2021. doi:10.1016/j.dcan.2020.12.006.
33. J. K. P. Seng, K. Li-minn Ang, E. Peter, and A. Mmonyi, "Artificial intelligence (AI) and machine learning for multimedia and edge information processing," *Electronics*, vol. 11, no. 14, p. 2239, 2022.
34. G. Kougioumtzidis, V. Poulkov, Z. D. Zaharis, and P. I. Lazaridis, "A Survey on Multimedia Services QoE Assessment and Machine Learning-Based Prediction," *IEEE Access*, vol. 10, pp. 19507–19538, 2022. doi:10.1109/ACCESS.2022.3149592.
35. A. Godinho, J. Rosado, F. S´a, F. Caldeira, and F. Cardoso, "Torrent Poisoning Protection with a Reverse Proxy Server," *Electronics*, vol. 12, no. 1, p. 165, 2023.
36. I. Shah, "Applications of Cloud Computing: A Review," *Authorea Preprints*, 2023.
37. P. Aneja, A. Bhatia, and A. Shankar, "A Review of Secure Cloud Storage-Based on Cloud Computing," in *Emerging Technologies in Data Mining and Information Security: Proceedings of IEMIS 2020, Volume 1*, pp. 923–933. Springer, 2021.
38. L. Bock, *Learn Wireshark: A definitive guide to expertly analyzing protocols and troubleshooting networks using Wireshark*. Packt Publishing Ltd, 2022.
39. A. Espinal, R. Estrada, and C. Monsalve, "Traffic model using a novel sniffer that ensures the user data privacy," in *MATEC Web of Conferences*, vol. 292, p. 03002. EDP Sciences, 2019.